\documentclass[prl,showpacs,twocolumn]{revtex4}

\usepackage{bm}
\usepackage{pifont}
\usepackage{amsmath,amsthm,cases}
\usepackage{ulem} 
\usepackage[dvipdfmx]{color}
\usepackage[dvipdfmx]{graphicx}

\begin{document}

\title{Microwave directional dichroism resonant with spin excitations in the polar ferromagnet GaV$_4$S$_8$}

\author{Y. Okamura$^1$}
\author{S. Seki$^2$}
\author{S. Bord$\rm{\acute{a}}$cs$^{3,4}$}
\author{$\rm{\acute{A}}$. Butykai$^3$}
\author{V. Tsurkan$^5$}
\author{I. K$\rm{\acute{e}}$zsm$\rm{\acute{a}}$rki$^{3}$}
\author{Y. Tokura$^{1,2}$}

\affiliation{
$^1$Department of Applied Physics and Quantum Phase Electronics Center, University of Tokyo, Tokyo 113-8656, Japan\\
$^2$RIKEN Center for Emergent Matter Science (CEMS), Wako 351-0198, Japan\\
$^3$Department of Physics, Budapest University of Technology and Economics and MTA-BME Lendulet Magneto-optical Spectroscopy Research Group, 1111 Budapest, Hungary\\
$^4$Hungarian Academy of Sciences, Premium Postdoctor Program, 1051 Budapest, Hungary\\
$^5$Experimental Physics V, Center for Electronic Correlations and Magnetism, University of Augsburg, 86159 Augsburg, Germany\\}

\begin{abstract}
We have investigated the directional dichroism of magnetic resonance spectra in the polar ferromagnet GaV$_4$S$_8$. While four types of structural domains are energetically degenerated under zero field, the magnetic resonance for each domain is well separated by applying magnetic fields due to uniaxial magnetic anisotropy. Consequently, the directional dichroism as large as 20 $\%$ is clearly observed without domain cancellation. The present observation therefore demonstrates that not only magnetoelectric mono-domain crystals but also magnetoelectric multi-domain specimens can be used to realize microwave (optical) diodes owing to the lack of inversion domains.
\end{abstract}

\maketitle

The light-matter interaction plays a fundamental role in various fields of physics, chemistry and engineering. The optical response is highly sensitive to the material properties even beyond the static ones and often help to realize versatile functionalities. Multiferroics, materials with simultaneously broken spatial inversion and time reversal symmetry, provide a unique arena to study novel optical phenomena that cannot show up in conventional ferromagnets or ferroelectrics \cite{kimura_nature2003,eerenstein_nature2006,cheong_natmat2007,tokura_rpp2014}. Such unconventional optical response, called the optical magnetoelectric (ME) effect, occurs when coexisting magnetic and ferroelectric orders simultaneously interact with the oscillating electric field and magnetic field of light. The discovery of this phenomenon was first reported for a noncentrosymmetric antiferromagnet Cr$_2$O$_3$ \cite{pisarev_phasetransit1991}, where the polarization rotation of light is reversed depending on the propagation direction of the light, although the effect is rather small. Another type of the optical ME effect, termed as nonreciprocal directional dichroism (DD), was demonstrated subsequently, where the light beams travelling in opposite directions are absorbed differently \cite{rikken_nature1997, kubota_prl2004,saito_jpsj2007,takahashi_natphys2011,kezsmarki_prl2011,bordacs_natphys2012,takahashi_prl2013,kezsmarki_natcom2014,bordacs_prb2015,kezsmarki_prl2015,yu_prl2017,okamura_natcom2013,okamura_prl2015,nii_jpsj2017,iguchi_natcom2017,kocsis_arxiv}. It has been revealed that the DD can become large and even show a cloaking function, thus being particularly important from the view point of future applications \cite{kezsmarki_natcom2014,kocsis_arxiv}.

The DD is observed for various types of excitations from X-ray to microwave-frequency regions in multiferroic materials \cite{rikken_nature1997, kubota_prl2004,saito_jpsj2007,takahashi_natphys2011,kezsmarki_prl2011,bordacs_natphys2012,takahashi_prl2013,kezsmarki_natcom2014,bordacs_prb2015,kezsmarki_prl2015,yu_prl2017,okamura_natcom2013,okamura_prl2015,nii_jpsj2017,iguchi_natcom2017,kocsis_arxiv}. Among them, the DD can often be large for the collective spin excitations that are both electric- and magnetic-dipole active owing to the strong ME coupling. These novel elementary excitations, referred to as the magnetoelectric resonances, are accompanied by the resonant motion of the magnetization ($\bm{M}$) as well as that of the electric polarization ($\bm{P}$), which enhances the ME coupling and leads to large DD. Recently, as a new guiding principle towards large DD, magnetoelectric resonances in the type-I multiferroics, in which magnetic orders develop within a pre-existing ferroelectric or pyroelectric phase \cite{khomskii_physics2009}, is attracting growing interests because this class of materials often exhibits large spin-induced changes of the electric polarization \cite{bordacs_prb2015,kezsmarki_prl2015,yu_prl2017}. The lacunar spinel GaV$_4$S$_8$ is a new member of the type-I multiferroics, which is an excellent candidate material to show large DD since it is a rare polar ferromagnet \cite{kezsmarki_natmat2015,ruff_sciadv2017}. 

In this Letter, we have studied the microwave DD at the ferromagnetic resonance in the polar magnetic semiconductor GaV$_4$S$_8$. While the sample consists of multiple polar rhombohedral structural domains in which the magnetic states are degenerate in zero magnetic field, the degeneracy of the magnetic excitations on the different types of domains is split in external magnetic fields owing to the axial magnetic anisotropy. As the result, the DD as large as 20 $\%$ is clearly observed free from cancelation among the domains. The magnitude critically depends on each domain, indicating the vital role of the microscopic ME coupling besides macroscopic lifted symmetry. 

\begin{figure}[htbp]
\begin{center}
\includegraphics[width=89mm]{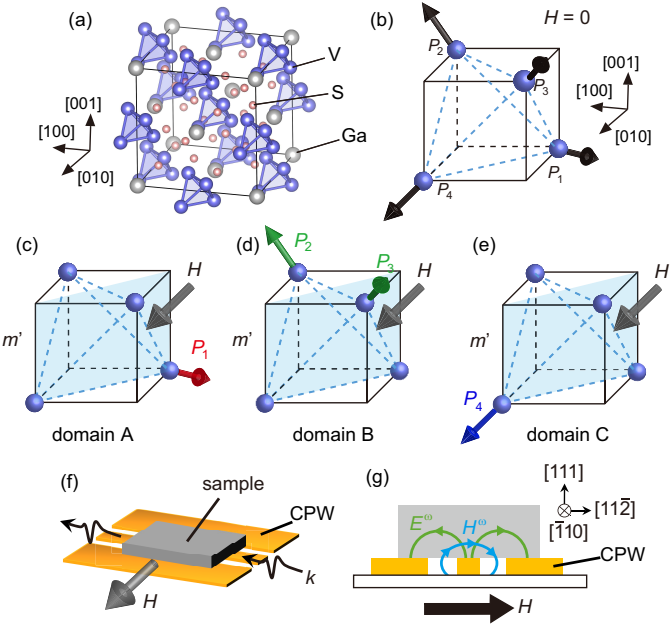}
\end{center}
\caption{(color online) (a) Crystal structure of GaV$_4$S$_8$ at room temperature. (b) The schematic illustration of V$_4$ clusters. The V ion is displaced along four equivalent cubic $\left<111\right>$ axes below the structural phase transition temperature. (c-e) Three types of structural domains classified based on the relationship with respect to $H\|[11\bar{2}]$. $m'$ represents a mirror plane combined with time reversal operation as indicated by the blue shaded region. (f) Schematic illustration of the coplanar wave guide (CPW). (g) Cross sectional view of the CPW; the spatial distribution of microwave electric and magnetic fields are illustrated.}
\end{figure}

The lacunar spinel structure of GaV$_4$S$_8$, which has the space group symmetry F$\bar{4}$3m at room temperature, is shown in Fig.~1(a) \cite{powell_chemmat2007,poch_chemmat2000}. It consists of a network of (V$_4$S$_4$)$^{5+}$ clusters that form a face centered cubic lattice. Below 42 K, due to cooperative Jahn-Teller distortion, the triply degenerate molecular orbitals of V$_4$ clusters are lifted through the elongation of the lattice along any of the four cubic $\left<111\right>$ axes [Fig.~1(b)]. The crystal structure then becomes polar (R3m) with a relatively large pyroelectric polarization, $P \simeq 1 \mu$C/cm$^2$ \cite{ruff_sciadv2017}. It should be noted that the opposite-polarization domains may exist due to the non-centrosymmetric nature of the room temperature cubic F$\bar{4}$3m phase. However, the previous PFM and static pyro/magnetocurrent measurements indicate that such inversion domains do not exist in these single crystals \cite{ruff_sciadv2017}. Therefore, we analyze the observed DD without considering the coexistence of inversion domains, which would lead to partial cancellation of the DD, i.e. its magnitude characteristic to a single inversion domain would be larger than the observed one. The magnetic transition occurs at $T_{\rm{c}}\sim$ 12.7 K, well below the structural transition temperature. The cycloidal spin state is stabilized at zero field between 6 $-$ 12.7 K and turns into a collinear field-polarized ferromagnetic state in moderate magnetic fields ($\bm{H}$) \cite{white_prb2018}. The N$\rm{\acute{e}}$el-type skyrmion lattice is observed in a specific temperature and magnetic field region \cite{kezsmarki_natmat2015,white_prb2018}. The stability of each magnetic phase critically depends on the magnetic-field direction due to the easy-axis anisotropy with respect to the rhombohedral axis, as discussed later [Figs.~1(c-e)].

Single crystals of GaV$_4$S$_8$ were grown with chemical vapor transport and the detail of the growth procedure is described elsewhere \cite{kezsmarki_natmat2015}. We performed broadband microwave spectroscopy to measure the transmission coefficient of the sample mounted on a coplanar waveguide (CPW) as shown in Fig.~1(f). The signal line was designed to be 20 $\mu$m in width so that it is much smaller than the sample width; the sample dimension is typically $\sim$ 2$\times$2$\times$1 mm$^3$. The directions of oscillating magnetic and electric field of the microwave, denoted as $\bm{H}^{\omega}$ and $\bm{E}^{\omega}$ respectively, depend on the position in space: the $\bm{H}^{\omega}$ just above the center of signal line is parallel to the plane of the CPW, while the $\bm{H}^{\omega}$ between the signal line and the ground is perpendicular to the plane [Fig.~1(g)]. The transmission coefficient for a microwave propagating along $\bm{k}^{\omega}$ and $-\bm{k}^{\omega}$ directions is denoted as $S_{12}$ and $S_{21}$, respectively, and was recorded with a vector network analyzer (Agilent Technology, E8363C). The absorption spectrum associated with the magnetic excitations, denoted as $\Delta S_{12}$ (or $\Delta S_{21}$) for the $+\bm{k}^{\omega}$ (or $-\bm{k}^{\omega}$) microwave, was obtained by calculating $\Delta S_{12(21)} =-S_{12(21)}+ S_{12(21)} (T > T_{\rm{c}})$, where $S_{12(21)}(T > T_{\rm{c}})$ taken at $T > T_{\rm{c}}$ does not contain the magnetic signal (0.01~$-$~35 GHz) (for more details, see Supplemental Material \cite{supple}).

\begin{figure}[htbp]
\begin{center}
\includegraphics[width=89mm]{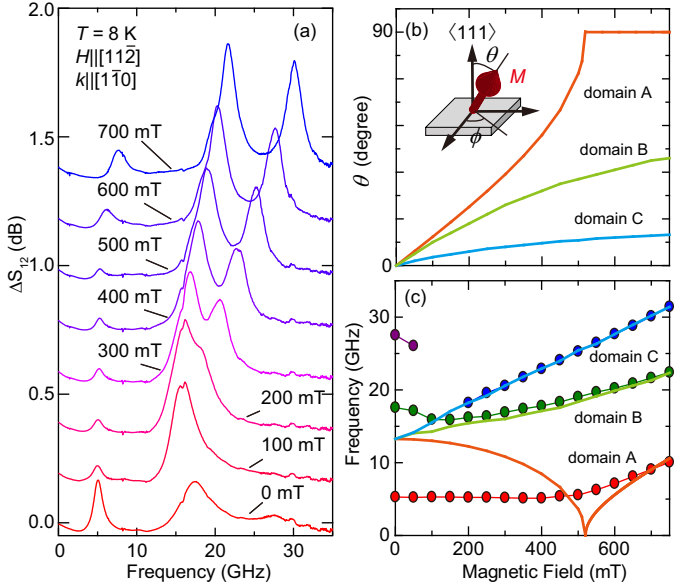}
\end{center}
\caption{(color online) (a) The magnetic resonance spectra for each magnetic field when $\bm{k}^{\omega}\|[1{\bar{1}}0]$ and $H\|[11{\bar{2}}]$. The spectra are shifted vertically for clarity. (b) Calculated angle of $\theta$, i.e., the angle between the magnetization direction and polar $\left<111\right>$ axis, numerically calculated based on Eq.~(1). (c) The magnetic-field dependence of the resonance frequency (circles). The calculated resonance frequency of ferromagnetic states (solid lines).}
\end{figure}

To observe the DD in this material, we focus on the magnetic resonance when $\bm{k}^{\omega}\|[1{\bar{1}}0]$ and $\bm{H}\|[11{\bar{2}}]$, as sketched in Fig.~1(g), because the DD can emerge for every types of structural domains from the symmetry point of view, as shown in Figs.~1(c-e). For this choice of the magnetic field, the single mirror plane remains for domain A and C. In contrast, two kinds of domain Bs are interchanged by this mirror reflection combined with time reversal operation. This unique mirror plane combined with the time reversal operation allows the emergence of DD for light beams propagating perpendicular to it but not for beams travelling parallel to the plane \cite{szaller_prb2013}. This symmetry is also compatible with a phenomenological toroidal moment $\bm{T}=\bm{P}\times\bm{M}$ pointing perpendicular to the mirror plane \cite{spaldin_jphys2008}. In all the three kinds of magnetic domains classified in terms of the angle between their $\left<111\right>$-type easy axes and the direction of the magnetic field (90 deg, 61.9 deg and 19.5 deg), $\bm{T}$ is finite, although it has different magnitudes in the different domains [Figs.~1(c-e) and Figs.~4(b,c)].

\begin{figure}[htbp]
\begin{center}
\includegraphics[width=89mm]{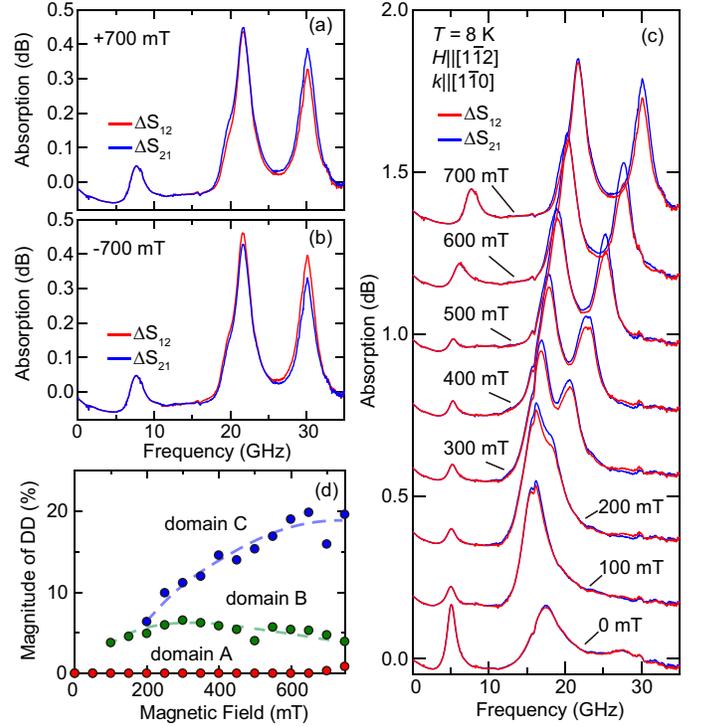}
\end{center}
\caption{(color online) The magnetic resonance spectra for $+\bm{k}^{\omega}$ and $-\bm{k}^{\omega}$ directions at +700 mT (a) and -700 mT (b). (c) The $+\bm{k}^{\omega}$ and $-\bm{k}^{\omega}$ spectra for each magnetic field. The spectra are shifted vertically for clarity. (d) The magnetic-field dependence of the magnitude of the directional dichroism. The red, green and blue circles correspond to the lower-, intermediate-, and higher-frequency modes, respectively.}
\end{figure}

As revealed in the previous magnetic-resonance study, the resonance frequency critically depends on the magnetic domains, or equivalently, the magnetic-field direction due to the uniaxial anisotropy \cite{ehlers_prb2016,ehlers_jphys2017}. Thus, we first investigated the evolution of the magnetic resonance with field for the different magnetic domains. Figure 2(a) shows the magnetic resonance spectra at 8 K for $\bm{k}^{\omega}\|[1{\bar{1}}0]$ in various magnetic fields $\bm{H}\|[11{\bar{2}}]$. In the cycloidal phase at zero field, three resonance peaks are observed at 5, 18 and 27 GHz, in accord with a previous study \cite{ehlers_prb2016}. By applying the magnetic field of 100 mT, the intensity of resonance modes at 5 and 18 GHz are weakened and enhanced, respectively, while the resonance peak discerned at 27 GHz immediately disappears. The absorption peak observed at 27 GHz, indicated as the purple dots in Fig.~2(c), is therefore the resonance mode characteristic of the cycloidal spin structure. With further increasing the field, the lower-lying mode slightly shifts towards higher-frequency region and the higher-lying mode splits into two modes. In domains B and C the cycloidal phase is suppressed and the field-polarized ferromagnetic state is reached by fields below $\sim$ 100 mT. In contrast, in domain A the transverse conical state formed in finite fields is robust to $\sim$ 500 mT, around which a spin-flop transition occurs to the ferromagnetic state. In the following we focus on the  magnetic resonances and their DD in the ferromagnetic state, where a simple phenomenological model can be used to describe the observed DD. 

In general, the resonance frequency of the ferromagnetic state can be calculated on the basis of the Smith-Suhl formula \cite{magosc}, $\left(\frac{\omega}{\gamma}\right)^2=\frac{1}{M^2 {\rm{sin}}^2 \theta} \left(\frac{\partial^2 E}{\partial \theta^2}  \frac{\partial^2 E}{\partial \phi^2} -\left(\frac{\partial^2 E}{\partial \theta \partial \phi }\right)^2 \right)$, where $\gamma$, $M$ and $E$ are the gyromagnetic ratio, magnetization and free energy, respectively. $\theta$ and $\phi$ are the polar angle and azimuthal angle of the $\bm{M}$, respectively, in a spherical coordinate system [Fig.~2(b), inset]. In the present system, the free energy is given by,
\begin{equation}
E=-\bm{M}\cdot\bm{H}-K\left(\bm{M}\cdot \bm{z}/M \right)^2,
\end{equation}
where $K$ and $\bm{z}$ represent the uniaxial anisotropy and unit vector along [111] axis, respectively. The resonance frequency is thus calculated as,
\begin{equation}
\frac{\omega}{\gamma}=\sqrt{H^2{\rm{sin}}^2\theta_H+\frac{H^2{\rm{cos}}\theta{\rm{sin}}2\theta_H}{2{\rm{sin}}\theta}+\frac{2KH{\rm{sin}}\theta_H{\rm{cos}}2\theta}{M{\rm{sin}}\theta}},
\end{equation}
where $\theta_H$ is the polar angle of the $\bm{H}$ from $\bm{z}$ axis. On the basis of Eq.~(1), $\theta$ can be numerically calculated for each domain so as to minimize $E$, as shown in Fig.~2(b). By substituting this result into Eq.~(2), it is found that the theory well reproduces the field dependence of the resonances in the ferromagnetic state with $K/M$ = 260 mT [Fig.~2(c)]. Here we used $g$ = 1.82 estimated in Ref.~\cite{ehlers_jphys2017} and the deduced $K/M$ value almost coincides with the value determined in Ref.~\cite{ehlers_jphys2017}. Thus, the low-, intermediate- and high-frequency modes observed at 700 mT are attributed to resonance modes in the domain A, B and C, respectively. Note that the discrepancy between the observed and calculated results for the domain C in lower field region is due to the subsisting transverse conical state up to 500 mT, not the ferromagnetic state assumed in the calculation.

Figures 3(a) and 3(b) show the absorption spectra when the microwave propagates along $+\bm{k}^{\omega}$ and $-\bm{k}^{\omega}$ directions at the field of $\pm$700 mT, where the magnetic state is ferromagnetic for all the domains. We found a clear signature of the DD for the intermediate- and high-frequency modes: The resonance peaks are higher in the $\Delta S_{21}$ spectrum than that in the $\Delta S_{12}$ spectrum at +700 mT for both modes. This relationship is reversed in $-$700 mT, further verifying the nonreciprocal nature of the transmission. The DD spectra change systematically by applying magnetic fields, as shown in Fig.~3(c). The magnitude of the DD, which is defined as $\frac{\Delta S_{12}-\Delta S_{21}}{(\Delta S_{12}+\Delta S_{21})/2}$, increases monotonically for the higher-lying mode and is basically unchanged for the intermediate mode as increasing the fields [Fig.~3(d)]. The magnitude of the DD amounts to approximately 20 $\%$ at +750 mT, which is the largest value among those reported for multiferroics in the microwave frequency range to the best of our knowledge \cite{okamura_natcom2013,okamura_prl2015,nii_jpsj2017,iguchi_natcom2017}. 

\begin{figure}[htbp]
\begin{center}
\includegraphics[width=89mm]{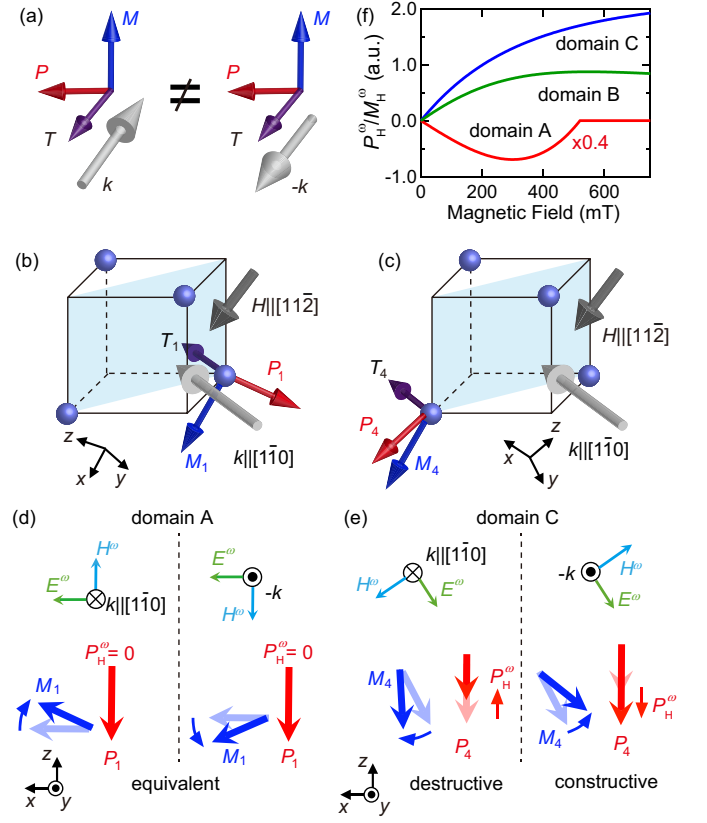}
\end{center}
\caption{ (color online) (a) Schematic illustration of the DD and toroidal moment. (b,c) The experimental configurations for domain A (b) and domain C (c). (d,e) Instantaneous response of $\bm{P}$ and $\bm{M}$ for the $P_1$ domain (d) and $P_4$ domain (e). (f) The magnetic-field dependence of the dynamical polarization induced by the microwave magnetic field in domain A, B and C.}
\end{figure}

Notably, although $\bm{k}^{\omega}$, $\bm{P}$ and $\bm{M}$ are perpendicular to each other in the domain A satisfying the necessary condition to observe the DD \cite{rikken_nature1997,kubota_prl2004,saito_jpsj2007,takahashi_natphys2011,kezsmarki_prl2011,takahashi_prl2013,kezsmarki_prl2015,yu_prl2017,okamura_natcom2013,iguchi_natcom2017}, the DD in the domain A is negligibly small. On the other hand, the domain C exhibits a strong DD, although for this type of domain $\bm{P}$ and $\bm{M}$ are nearly antiparallel, i.e. toroidal moment $\bm{T}=\bm{P}\times\bm{M}$ is smaller as compared to the domain A [Figs.~4(b-c)]. Moreover, the domains A and C are related with each other by rotating 70.5 degrees around the $\bm{k}^{\omega} \|[1\bar{1}0]$. These facts highlight the crucial role the direction of the $\bm{M}$ in the ME coupling. It should be noted that, although the Damon-Eshbach mode or magnetostatic surface wave (at non-zero wavenumber) may show the nonreciprocal propagation, it does not lead to the nonreciprocal absorption \cite{magosc}. Moreover, the Damon-Eshbach mode cannot be clearly discerned in the present experiment, thus allowing us to exclude its contribution to the presently observed DD. 

To understand the role of the ME coupling, we consider the instantaneous responses induced by $\bm{H}^{\omega}$ and $\bm{E}^{\omega}$ in the same manner as adopted in Ref.~\cite{okamura_prl2015}. From the view point of the symmetry, the $\bm{P}$ can be described by,
\begin{equation}
\begin{split}
P_x=aM_xM_z-bM_xM_y,\\
P_y=-bM_x^2+bM_y^2+aM_yM_z, \\
P_z=cM_z^2+dM^2,
\end{split}
\end{equation}
where $x$, $y$ and $z$ represent $[11{\bar{2}}]$, $[{\bar{1}}10]$ and [111] axes, respectively. The terms that contain $H$ such as $P_x=a'H_xH_z$ are also allowed and have the same form as $M$ but we omit them here for clarity. Given that the $+\bm{k}^{\omega}(\|-y)$ microwave in the CPW contains $\bm{H}^{\omega}\|z$ and $\bm{E}^{\omega}\|x$ [Fig.~1(g)] and that the $\bm{M}$ points to the $x$ direction in the domain A with $P_1$, the magnetic resonance is induced by $\bm{H}^{\omega}\|z$ and produces the dynamical magnetization, $\bm{M}_H^{\omega}$, along the $z$ direction [Fig.~4(d)]. $\bm{M}_H^{\omega}$ simultaneously induces dynamical polarization, $\bm{P}_H^{\omega}$, along the $x$ direction through the ME coupling given by Eq.~(3); $P_{H,x}^{\omega} = a M_x M^{\omega}_{H,z}$. This $P_{H,x}^{\omega}$ may interfere with the dynamical polarization ($\bm{P}_E^{\omega}$) induced by $\bm{E}^{\omega} (\|x)$ and lead to the DD. In the present case, however, the DD is experimentally not observed in the domain A and therein the coefficient $a$ should be negligibly small.

In the domain C with $P_4$, the $\bm{M}$ is slanted from both $x$ and $z$ axes and therefore $\bm{M}$ and $\bm{M}^{\omega}_H$ have both finite $x$ and $z$ components [Fig.~4(e)]. As the result, $\bm{P}_H^{\omega}$ is generated along $z$ direction, which can be described as $2cM_z M_{H,z}^{\omega}$. The resulting $P_{H,z}^{\omega}$ is antiparallel to $P^{\omega}_{E,z}$ for $+\bm{k}^{\omega}$ microwave whereas they are parallel for $-\bm{k}^{\omega}$ microwave, as schematically illustrated in Fig.~4(d); here, we note that $\bm{H}^{\omega}$ and $\bm{E}^{\omega}$ also have both $x$ and $z$ components [Fig.~1(d)]. Therefore, the interference between $P_{H,z}^{\omega}$ and $P^{\omega}_{E,z}$ occurs destructively for $+\bm{k}^{\omega}$ microwave while constructively for $-\bm{k}^{\omega}$ microwave. This difference may lead to the DD, which is indeed observed in Fig.~3. Hence, the DD observed here originates from the form of the ME coupling, $P_z=cM_z^2$, which is in accord with the nature of the static ME effect reported in Ref.~\cite{ruff_sciadv2017}. The microscopic origin of the ME coupling is therefore attributed either to anisotropic exchange mechanism or to single-site ME effect.

The ME coupling discussed above reproduces the magnetic-field dependence of the DD in the ferromagnetic state as well: Because the $\bm{H}^{\omega}$ circulates around the signal line and is perpendicular to the [110] axis in the present experimental setup, we calculated $P_{H,z}^{\omega} = 2c M_z M_{H,z}^{\omega}$, which is proportional to the magnitude of the DD, for each domain by averaging all the contributions induced by $\bm{H}^{\omega}$ lying in the (110) plane. The calculation qualitatively reproduces the experiment except for the low-field region for domain A, where the transverse conical state, not the ferromagnetic state presumed in the calculation, is formed: The DD for domain A is negligibly small, for domain B shows a broad maximum and for domain C keeps increasing with the magnetic field (see Figs.~3(d) and 4(f)).
 
It has been believed that the DD should be maximized when $\bm{k}^{\omega}$, $\bm{P}$ and $\bm{M}$ are perpendicular to each other, whereas the present observation unambiguously demonstrates the crucial role of the microscopic ME coupling as well as the macroscopic symmetry: The macroscopic symmetry defines the necessary conditions for the existence of the DD, as illustrated in Fig.~4(a), but leaves freedom for different ME mechanisms to govern the magnitude of the DD; as clear from the comparison of the magnetization and polarization dynamics for domain A and C, the DD can emerge even when $\bm{k}^{\omega}$, $\bm{P}$ and $\bm{M}$ are not totally perpendicular to each other.

In summary, we have investigated the microwave DD in the polar ferromagnetic state of GaV$_4$S$_8$. The ferromagnetic resonance is separated for the structural domains with different directions of the electric polarization due to uniaxial anisotropy and we clearly observed the DD as large as 20 $\%$ for a specific domain without cancelation among the multi-domain states. Our findings widen the class of materials that can potentially show the large DD even in the presence of the multi-domain states.

The authors thank M. Mochizuki, K. Penc and T. Kurumaji for enlightening discussions. This work was supported by the Grant-in-Aid for Scientific Research (Grant Nos. 24224009 and 24226002) from the JSPS, Murata Science foundation, the Hungarian National Research, Development, the BME-Nanonotechnology and Materials Science FIKP grant of EMMI (BME FIKP-NAT), and Innovation Office-NKFIH via Grant No. ANN 122879, the Deutsche Forschungsgemeinschaft (DFG) via the Transregional Research Collaboration TRR 80: From Electronic Correlations to Functionality (Augsburg-Munich-Stuttgart) and via the Skyrmionincs Priority Program SPP2137.

\clearpage


\begin{thebibliography}
-
\bibitem{kimura_nature2003} T. Kimura, T. Goto, H. Shintani, K. Ishizaka, T. Arima, and Y. Tokura, Nature {\bf{426,}} 55 (2003).
\bibitem{eerenstein_nature2006} W. Eerenstein, N. D. Mathur, and J. F. Scott, Nature {\bf{442,}} 759 (2006).
\bibitem{cheong_natmat2007} S.-W. Cheong, and M. Mostovoy, Nat. Mater. {\bf{6,}} 13 (2007).
\bibitem{tokura_rpp2014} Y. Tokura, S. Seki, and N. Nagaosa, Rep. Prog. Phys. {\bf{77,}} 076501 (2014).
\bibitem{pisarev_phasetransit1991} R. V. Pisarev, B. B. Krichevtsov, and V. V. Pavlov, Phase Transit. {\bf{37,}} 63 (1991).
\bibitem{rikken_nature1997} G. L. Rikken, and E. Raupach, Nature {\bf{390,}} 493 (1997).
\bibitem{kubota_prl2004} M. Kubota, T. Arima, Y. Kaneko, J. P. He, X. Z. Yu, and Y. Tokura, Phys. Rev. Lett. {\bf{92,}} 137401 (2004).
\bibitem{saito_jpsj2007} M. Saito, K. Taniguchi, and T. Arima, J. Phys. Soc. Jpn. {\bf{77,}} 013705 (2007).
\bibitem{kezsmarki_prl2011} I. K$\rm{\acute{e}}$zsm$\rm{\acute{a}}$rki, N. Kida, H. Murakawa, S. B$\rm{\acute{o}}$rdacs, Y. Onose, and Y. Tokura, Phys. Rev. Lett. {\bf{106,}} 057403 (2011).
\bibitem{takahashi_natphys2011} Y. Takahashi, R. Shimano, Y. Kaneko, H. Murakawa, and Y. Tokura, Nat. Phys. {\bf{8,}} 121 (2011).
\bibitem{bordacs_natphys2012} S. Bord$\rm{\acute{a}}$cs, I. K$\rm{\acute{e}}$zsm$\rm{\acute{a}}$rki, D. Szallar, L. Demk$\rm{\acute{o}}$, N. Kida, H. Murakawa, Y. Onose, R. Shimano, T. R${\rm \tilde{o} \tilde{o}}$m, U. Nagel, S. Miyahara, N. Furukawa, and Y. Tokura, Nat. Phys. {\bf{8,}} 734 (2012).
\bibitem{takahashi_prl2013} Y. Takahashi, Y. Yamasaki, and Y. Tokura {\bf{111,}} 037294 (2013).
\bibitem{kezsmarki_natcom2014} I. K$\rm{\acute{e}}$zsm$\rm{\acute{a}}$rki, N. Szaller, S. Bord$\rm{\acute{a}}$cs, H. Murakawa, Y. Tokura, H. Engelkamp, T. R${\rm \tilde{o} \tilde{o}}$m, and U. Nagel, Nat. Commun. {\bf{5,}} 3203 (2014).
\bibitem{kocsis_arxiv} V. Kocsis, K. Penc, T. R${\rm \tilde{o} \tilde{o}}$m, U. Nagel, J. V$\rm{\acute{i}}$t, Y. Tokunaga, Y. Taguchi, Y. Tokura, I. K$\rm{\acute{e}}$zsm$\rm{\acute{a}}$rki, and S. Bord$\rm{\acute{a}}$cs, Phys. Rev. Lett. {\bf{121,}} 057601 (2018).
\bibitem{bordacs_prb2015} S. Bord$\rm{\acute{a}}$cs, V. Kocsis, Y. Tokunaga, U. Nagel, T. R${\rm \tilde{o} \tilde{o}}$m, Y. Takahashi, Y. Taguchi, and Y. Tokura, Phys. Rev. B {\bf{92,}} 214441 (2015).
\bibitem{kezsmarki_prl2015} I. K$\rm{\acute{e}}$zsm$\rm{\acute{a}}$rki, U. Nagel, S. Bord$\rm{\acute{a}}$cs, R. S. Fisherman, J.H. Lee, Hee Taek Yi, S.-W. Cheong, and T.R${\rm \tilde{o} \tilde{o}}$m, Phys. Rev. Lett. {\bf{115,}} 127203 (2015).
\bibitem{yu_prl2017} S. Yu, B. Gao, J.W. Kim, S-W. Cheong, M.K.L. Man, J. Mad$\rm{\acute{e}}$o, K.M. Dani, and D. Talbayev, Phys. Rev. Lett. {\bf{120,}} 037601 (2017).
\bibitem{okamura_natcom2013} Y. Okamura, F. Kagawa, M. Mochizuki, M. Kubota, S. Seki, S. Ishiwata, M. Kawasaki, Y. Onose, and Y. Tokura, Nat. Commun. {\bf{4,}} 2391 (2013).
\bibitem{okamura_prl2015} Y. Okamura, F. Kagawa, S. Seki, M. Kubota, M. Kawasaki, and Y. Tokura, Phys. Rev. Lett. {\bf{114,}} 197202 (2015).
\bibitem{nii_jpsj2017} Y. Nii, R. Sasaki, Y. Iguchi, and Y. Onose, J. Phys. Soc. Jpn. {\bf{86,}} 024707 (2017).
\bibitem{iguchi_natcom2017} Y. Iguchi, Y. Nii, and Y. Onose, Nat. Commun. {\bf{8,}} 15252 (2017).
\bibitem{khomskii_physics2009} D. Khomskii, Physics {\bf{2,}} 20 (2009).
\bibitem{kezsmarki_natmat2015} I. K$\rm{\acute{e}}$zsm$\rm{\acute{a}}$rki, S. Bord$\rm{\acute{a}}$cs, P. Milde, E. Neuber, L. M. Eng, J. S. White, H. M. R$\o$nnow, C. D. Dewhurst, M. Mochizuki, K. Yanai, H. Nakamura, D. Ehlers, V. Tsurkan, and A. Loidl, Nat. Mater. {\bf{14,}} 1116 (2015).
\bibitem{ruff_sciadv2017} E. Ruff, S. Widmann, P. Lunkenheimer, V. Tsurkan, I. K$\rm{\acute{e}}$zsm$\rm{\acute{a}}$rki, and A. Loidl, Sci. Adv. {\bf{1,}} e1500916 (2015).
\bibitem{butykai_scirep2017} $\rm{\acute{A}}$. Butykai, S. Bord$\rm{\acute{a}}$cs, I. K$\rm{\acute{e}}$zsm$\rm{\acute{a}}$rki, V. Tsurkan, A. Loidl, J. D$\rm{\ddot{o}}$ring, E. Neuber, P. Milde. S C. Kehr, and L. M. Eng, Sci. Rep. {\bf{7,}} 44663 (2017).
\bibitem{poch_chemmat2000} R. Pocha, D. Johrendt, and R. P$\rm{\ddot{o}}$ttgen, Chem. Mater. {\bf{12,}} 2882 (2000).
\bibitem{powell_chemmat2007} A. V. Powell, A. McDowall, I. Szkoda, K. S. Knight, and B. J. Kennedy, T. Vogt, Chem. Mater. {\bf{19,}} 5035 (2007).
\bibitem{white_prb2018} J. S. White, $\rm{\acute{A}}$. Butykai, R. Cubitt, D. Honecker, C. D. Dewhurst, L. F. Kiss, V. Tsurkan, and S. Bord$\rm{\acute{a}}$cs, Phys. Rev. B {\bf{97,}} 020401 (2018).
\bibitem{supple} The supplemental data and the analysis procedure are shown in Supplemental Material at http://link.aps.org/
supplemental/10.1103/PhysRevLett.xxx.
\bibitem{szaller_prb2013} D. Szaller, S. Bord$\rm{\acute{a}}$cs, and I. K$\rm{\acute{e}}$zsm$\rm{\acute{a}}$rki, Phys. Rev. B {\bf{87,}} 014421 (2013).
\bibitem{spaldin_jphys2008} N. A. Spaldin, M. Fiebig, and M. Mostovoy, J. Phys. Condens. Matter {\bf{20,}} 434203 (2008).
\bibitem{ehlers_prb2016}D. Ehlers, I. Stasinopoulous, V. Tsurkan, H.-A. Krug von Nidda, T. Feh$\rm{\acute{e}}$r, A. Leonov, I. K$\rm{\acute{e}}$zsm$\rm{\acute{a}}$rki, D. Grundler, and A. Loidl, Phys. Rev. B {\bf{94,}} 014406 (2016).
\bibitem{ehlers_jphys2017} D. Ehlers, I. Stasinopoulos, I. K$\rm{\acute{e}}$zsm$\rm{\acute{a}}$rki, T. Feh$\rm{\acute{e}}$r, V. Tsurkan, H-A. Krug von Nidda, D. Grundler and A. Loidl, J. Phys. Condens. Matter {\bf{29,}} 065803 (2017).
\bibitem{magosc} A. G. Gurevich and G. A. Melkov, Magnetization Oscillations and Waves (CRC Press, 1996).
\end{thebibliography}
\end{document}